\documentclass[journal,12pt,onecolumn,draftclsnofoot]{IEEEtran}

\usepackage{amsmath,amsfonts}
\usepackage{graphicx,amssymb}
\usepackage{algorithmic,algorithm}
\usepackage{amsthm,multirow}
\usepackage{subfigure,url,color,capt-of,booktabs,hhline}
\usepackage[noadjust]{cite}

\usepackage{cite}
\usepackage{graphics}

\ifCLASSINFOpdf
\else
\fi

\newcommand{\bG}{\mathbf{G}}
\newcommand{\bZ}{\mathbf{Z}}

\begin{document}
\title{Smart and Reconfigurable Wireless Communications: From IRS Modeling to Algorithm Design}

\author{Xianghao Yu, Vahid Jamali, Dongfang Xu, Derrick Wing Kwan Ng, and Robert Schober% <-this % stops a space
\thanks{Xianghao Yu is with the Hong Kong University of Science and Technology; Vahid Jamali, Dongfang Xu, and Robert Schober are with Friedrich-Alexander University Erlangen-Nuremberg; Derrick Wing Kwan Ng is with University of New South Wales.}}

\maketitle

\begin{abstract}
Intelligent reflecting surfaces (IRSs) have been introduced into wireless communications systems due to their great potential to smartly customize  and reconfigure radio propagation environments in a cost-effective manner.
Despite the promising advantages of IRSs,  academic research on IRSs is still in its infancy.
In particular, the design and analysis of IRS-assisted wireless communication systems critically depend on an accurate and tractable modeling of the IRS.
In this article, we first present and compare three  IRS models, namely the conventional independent diffusive scatterer-based model, physics-based model, and impedance network-based model, in terms of their  accuracy, tractability, and hardware complexity.  
Besides, a new framework based on partitioning the IRS into tiles and employing codebooks of transmission modes is introduced to enable scalable IRS optimization.
Then, we investigate the impact of the three considered IRS models on system design, where several crucial technical challenges for the efficient design of IRS-assisted wireless systems are identified and the corresponding solutions are unraveled.
Furthermore, to illustrate the properties of the considered models and the efficiency of the proposed solution concepts, IRS-assisted secure wireless systems and simultaneous wireless information and power transfer (SWIPT) systems are studied  in more detail.
Finally, several promising future research directions for IRS-assisted wireless systems are highlighted.

\end{abstract}

\begin{IEEEkeywords}
Algorithm design, intelligent reflecting surfaces, IRS modeling.
\end{IEEEkeywords}

\newpage

% \fi
\IEEEpeerreviewmaketitle

\section{Introduction}
%\IEEEPARstart{T}{his} 
In legacy wireless communications systems, wireless channels are typically considered to be uncontrollable and treated as ``black boxes''. Thus, various advanced communication techniques have been proposed to adapt to the given properties of these boxes.
Recently, reconfigurable intelligent surfaces (RISs) have stood out as a promising enabler to break this stereotype. 
In particular, as a kind of programmable metasurfaces, RISs are able to customize wireless signal propagation, which opens new avenues for realizing smart radio environments in future sixth-generation (6G) wireless systems \cite{8910627}.
Among a variety of RISs, intelligent reflecting surfaces (IRSs) have drawn special attention from both academia and industry due to their low power consumption and economical implementation cost.
Specifically, IRSs are typically implemented by a large number of \emph{passive} elements, e.g., diodes and phase shifters, and do not require active hardware components such as radio frequency (RF) chains \cite{8741198}. 
Thus, IRSs consume limited power for operation (each element consumes typically less than 1 mW), which aligns with the growing  need for green wireless communications \cite{emil}.
Furthermore, IRSs can be fabricated as artificial thin films that can be readily attached to the facades of infrastructures, e.g., high-rises and overpasses, which significantly reduces implementation complexity.

The benefits of IRSs have been confirmed for various wireless communication scenarios in recent literature, including physical layer security provisioning \cite{9133130}, full-duplex transmission \cite{9183907},  millimeter-wave wireless networks \cite{9322519}, and simultaneous wireless information and power transfer (SWIPT) systems \cite{WCNC}.
To fully unleash the potential of IRSs, they have to be carefully configured and their multifaceted impact on the performance of wireless systems has to be accurately characterized. However, these challenges have not been satisfactorily addressed, yet.
%However, efficient design and comprehensive understanding of the effect of IRSs in wireless systems 

A fundamental obstacle in this regard is the lack of well-balanced IRS models for both system optimization and performance evaluation of IRS-aided wireless  systems.
%In particular, the inactive hardware components at IRSs bring magnificent challenges in delicately optimizing and accurately analyzing IRS-assisted wireless systems \cite{8741198}.
%To fully unleash the potential of IRSs in wireless systems, they have to be delicately designed and analyzed together with conventional communication techniques.
%, which forms the first technical challenge in the design of IRS-assisted wireless systems.
%a well-rounded communication model for IRSs in wireless systems is desired for both system optimization and performance evaluation, which is a fundamental task in IRS-empowered wireless systems.
In particular, there  exists a trade-off among different priorities when modeling IRSs, i.e., accuracy, tractability, and hardware complexity.
More importantly, how the IRSs are modeled crucially impacts the principles and methodologies applicable for the design of IRS-aided wireless systems. 
So far, a systematic comparison between existing IRS models and their implications for wireless system design do not exist.

The goal of this article is to provide a comprehensive overview of different IRS models and to study their impact on the design of IRS-assisted wireless systems.
We investigate three existing IRS models in this article. The first model is the conventional IRS model that has been widely adopted in the literature \cite{8910627} while the other two have been recently proposed and address the need for more accurate physical propagation environment characterization \cite{najafi2020intelligent} and enhanced IRS capabilities \cite{shen2020modeling}, respectively.
In addition, a new framework is introduced for scalable IRS optimization.
Then, key challenges for the design of IRS-empowered wireless systems are identified, where potential technical solutions are discussed for the different considered IRS models.
To  provide a deeper understanding of the different IRS models and solution concepts, we elaborate on two specific application scenarios focusing on secure wireless communications and SWIPT systems. 
Furthermore, exciting open problems and future research directions are also highlighted.

\section{IRS Modeling}\label{Sec:IRSmodel}

In this section, we introduce three theoretical IRS models for wireless communications and present a framework for scalable IRS design,
c.f. Fig.~\ref{Fig:Models} and Table I.

%\newpage
\subsection{Independent Diffusive Scatterer-based (IDS) Model}\label{Sec:IDSmodel}

\begin{figure}[h]
	\centering
	\includegraphics[width=1\textwidth]{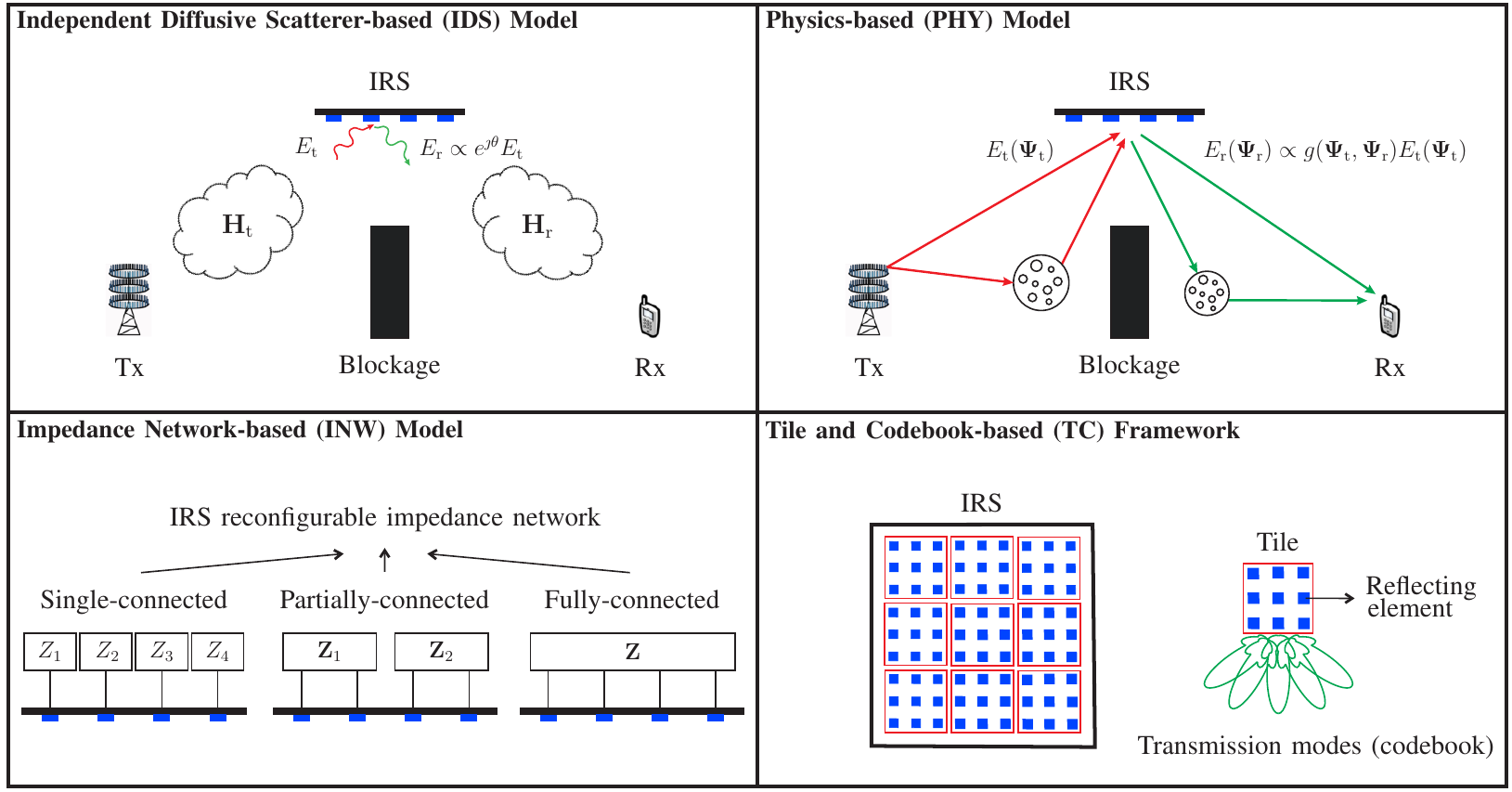} 	
	\caption{Illustration of different models and the TC framework discussed in this paper. For the IDS model, $E_\mathrm{t}$ and $E_\mathrm{r}$ denote the electric field on a given reflecting element,  respectively, whereas for the PHY model, $E_\mathrm{t}(\mathbf{\Psi}_\mathrm{t})$ and $E_\mathrm{r}(\mathbf{\Psi}_\mathrm{r})$ denote the electric field on the IRS for a wave coming from AoA $\mathbf{\Psi}_\mathrm{t}$ and  the electric field in the far-field of the IRS along AoD $\mathbf{\Psi}_\mathrm{r}$, respectively.  }
	\label{Fig:Models}			
\end{figure}

A widely-adopted model for IRSs in the literature of wireless communications  is to assume that each reflecting element individually acts as a diffusive scatterer that is able to alter the phase of the impinging electromagnetic (EM) wave during reflection \cite{8910627}.  
%Thereby, the IRS is modeled by the so-called phase shift matrix, denoted by $\bTheta$, which is a diagonal matrix with reflection coefficient $\beta_q \e^{\jj \theta_q}$ as its $q$-th diagonal entry. 
Thereby, the impact of the IRS is modeled by a diagonal matrix $\mathbf{\Phi}$, called phase shift matrix, whose non-zero entries are the reflection coefficients.
Since IRSs are  typically passive and to conserve the total energy during reflection, the magnitudes of the reflection coefficients are set  to  one, i.e., \emph{unit modulus} reflection coefficients.
%Here, $\theta_q\in[0,2\pi)$ denotes the phase shift applied by the $q$-th reflecting element and is optimized to realize different functionalities of the IRS. 
%Moreover, $\beta_q\in[0,1]$ denotes the amplitude of the reflection coefficient and is often assumed to be equal to one, i.e., passive loss-less IRS. 
%In the literature, the aforementioned IRS model is often jointly used with Tx-to-IRS and IRS-to-Rx channel matrices, denoted by $\bH_\mathrm{t}$ and $\bH_\mathrm{r}$, respectively, that are modeled with i.i.d. entries following Rayleigh or Rician distributions \cite{8910627}. 
Throughout this paper, we refer to this model as the IDS model and treat it as a baseline model for more sophisticated IRS models, see Fig.~\ref{Fig:Models}. 

While the IDS model accounts for the basic properties of IRSs, e.g., the phase shift introduced by each reflecting element and IRS passivity, it suffers from the following limitations.
\begin{itemize}
	\item The physical properties of IRSs, e.g., the size of the reflecting elements, polarization, connectivity  among reflecting elements, and wave angle-of-arrival (AoA) and angle-of-departure (AoD), are not \textit{explicitly} modeled. 
	Hence, IRS-assisted systems designed based on the IDS model cannot effectively leverage these important and practical properties. 
	%	\item The assumption of independent entries for $\bH_\mathrm{t}$ and $\bH_\mathrm{r}$ become invalid for \textit{large IRSs}.
	\item The unit modulus constraint on the reflection coefficients significantly complicates the resource allocation algorithm design \cite{9133130,9183907} making it \textit{not scalable} for large IRSs. 
\end{itemize}
Next, we discuss more elaborate IRS models that address the above challenges of the IDS model.

\subsection{Physics-based (PHY) Model}
\label{Sec:PHYmodel}

While research on modeling and analysis of intelligent surfaces has a rich history in the physics and electromagnetics literature, the development of EM-compliant IRS models from a communication-theoretical perspective has only recently attracted attention \cite{najafi2020intelligent,di2020smart}. 
For instance, in \cite{najafi2020intelligent}, the EM discontinuities imposed by the IRS were modeled by using effective surface currents and the reflected wave from the IRS was analyzed by solving Maxwell's equations for the electric and magnetic vector fields. 
Also, IRSs were modeled as  arrays of electrically and magnetically polarizable reflecting elements in \cite{di2020smart}. 
%In addition,  analytical expressions based on the generalized sheet transition conditions were presented to analyze the reflected wave.  
%Next, we discuss the deterministic path loss model and the statistical channel models  developed in \cite{najafi2020intelligent} in more details. 
Next, we discuss the main ideas of the proposed PHY model.

One key motivation of exploiting physical information for IRS modeling is to properly capture the unique radio propagation environment in IRS-assisted wireless systems.
In particular, the number of channel scatterers  in wireless systems is typically limited, especially when the direct link between the transceivers is blocked.
Hence, accurately reflecting the impinging EM waves to the directions that associate with strong paths in the channel is crucial for the IRS to enhance system performance.
Assuming a far-field scenario, an IRS can be modeled by the \emph{generalized radar cross section (GRCS)}, denoted by $g(\mathbf{\Psi}_\mathrm{t},\mathbf{\Psi}_\mathrm{r})$, which 
determines how a plane wave impinging from an AoA $\mathbf{\Psi}_\mathrm{t}$ with a given polarization is reflected in an intended AoD $\mathbf{\Psi}_\mathrm{r}$ for a given phase shift configuration of the IRS \cite{najafi2020intelligent}.
%It was shown in \cite[Lemma~1]{najafi2020intelligent} that the end-to-end IRS-assisted channel path loss can be decomposed into the product of the free-space path losses of the Tx-to-IRS channel and IRS-to-Rx channels, and a term proportional to generalized RCS of the IRS. 
%Using this model, the authors derived the required minimum  number of IRS reflecting elements, denoted by $Q$, to achieve a given link budget, cf. \cite[Corollary~4]{najafi2020intelligent}.  For example, for Tx-to-IRS and IRS-to-Rx distances of $100$~m,  half-wavelength unit-cell spacing, and  carrier frequency of $5$~GHz, it was shown that at least $Q\approx3300$ reflecting elements are needed such that the path loss of the IRS-assisted channel becomes equal to that of an unobstructed direct link with the same end-to-end distance of $200$~m.
%\textbf{Statistical IRS-assisted channel model:} 
%In \cite{najafi2020intelligent}, the Tx-to-IRS and IRS-to-Rx channels are modeled by a collection of channel scatterer clusters, which are characterized by their AoAs, AoDs, and effective channel gains,  see Fig.~\ref{Fig:Models}. The channel paths within each scatterer cluster are indistinguishable at the Tx, Rx, and IRS and their combined effect is modeled by random fading coefficient, cf. Table~I. 
Mathematically, one can adopt a GRCS matrix $\bG$, whose entries are $g(\mathbf{\Psi}_\mathrm{t},\mathbf{\Psi}_\mathrm{r})$  evaluated at different IRS AoAs and AoDs, to model the IRS. Note that in addition to the wave AoAs and AoDs, the IRS GRCS also accounts for other physical properties of the IRS such as the size of the reflecting elements and the distance between the reflecting elements, which are not taken into account in the IDS model.

%We refer to this model as physics-based (PHY) IRS-assisted channel model since it accounts for the physical properties of the IRS and the Tx-to-IRS and IRS-to-Rx links.

%It is argued in \cite{di2020smart} that the mutual coupling among reflecting elements can be \textit{implicitly} account for if the IRS is described by a \textit{surface-averaged} impedance (or susceptibility) function. However, obtaining these surface-averaged functions  is challenging as it require either full-wave simulation or solving integral equations. Therefore, in the following, we discuss a model that \textit{explicitly} accounts for the mutual coupling among reflecting elements. 

\subsection{Impedance Network-based (INW) Model}
\label{Sec:INWmodel}
%In particular, one may distinguish two impedance matrices, namely one that accounts for the mutual coupling among IRS reflecting elements over the air and the other that describes the control network connectivity of IRS reflecting elements, see Fig.~\ref{Fig:Models}.    

%\textbf{IRS mutual coupling:} The IRS reflecting elements are often separated  on the order of sub-wavelength apart, which pronounces their mutual coupling. The impact of mutual coupling can be modeled via the so-called \textit{mutual impedance matrix} $\bS$, see \cite{gradoni2021end} for physics-based characterization of $\bS$. 
%Alternatively, in \cite{shen2020modeling}, the IRS is modeled as a multi-port network specified by a general \textit{scattering matrix} that models the impact of mutual coupling.    

%\textbf{IRS control connectivity:} 

In the literature, it is often assumed that each IRS reflecting element is separately controlled by a tunable circuit which can be modeled as a tunable impedance.
For example, an impedance-based representation of the IDS model was provided in \cite{9115725}. 
In contrast, in \cite{shen2020modeling}, it was proposed to connect all or a subset of IRS reflecting elements via an impedance network and jointly control them via an effective impedance matrix, denoted by $\bZ$.
In this way, the entire IRS is modeled as a multi-port network characterized by a general \textit{scattering matrix} $\mathbf{\Theta}$.
Depending on how the reflecting elements are connected, IRSs can be categorized into the following three architectures, see also Fig.~\ref{Fig:Models}.
\begin{itemize}
	\item \textbf{Single-connected (SC) IRS:} For this architecture, the IRS reflecting elements are not connected to each other. In this case, the INW model reduces to the baseline IDS model, i.e., $\mathbf{\Theta=\Phi}$, and the corresponding impedance matrix $\mathbf{Z}$ is the same as the one presented in \cite{9115725}.
%	Hence, the scattering matrix $\mathbf{\Theta}$ reduces to the phase shift matrix $\mathbf{\Phi}$ of the IDS model.
	\item \textbf{Fully-connected (FC) IRS:} For this architecture, each IRS reflecting element is connected via an impedance to all other reflecting elements, which results in 
%	a full IRS impedance matrix $\bZ$ and, correspondingly, 
	 a complex symmetric unitary scattering matrix $\mathbf{\Theta}$ \cite{shen2020modeling}. 
	\item \textbf{Partially-connected (PC) IRS:} This architecture is a compromise between the previous two  where the IRS reflecting elements  are divided into groups and all reflecting elements within a group are fully connected. Correspondingly, the scattering matrix $\mathbf{\Theta}$ is a block diagonal matrix where each submatrix is a complex symmetric unitary matrix. 
\end{itemize}   

By connecting the reflecting elements, either fully or partially, via a configurable impedance network, the scattering matrix $\mathbf{\Theta}$ 
%is not restricted to a unit modulus diagonal matrix as $\mathbf{\Phi}$ is in the IDS model. Instead, thanks to the additional degrees of freedom provided by the impedance networks, $\mathbf{\Theta}$ 
is  composed of complex symmetric unitary submatrices, which constitutes a generalization of the unit modulus diagonal matrix $\mathbf{\Phi}$ in the IDS model.
This indicates that thanks to the additional degrees of freedom (DoFs) provided by the impedance network, the INW model is able to adjust not only the phases but also the magnitudes of  impinging waves, which can further enhance possible performance gains compared with the previous two IRS models not employing impedance networks.
In addition, the GRCS for the PHY model can also be integrated with the INW model, which not only improves performance but also makes the overall model physically explainable.

%In addition to manually controlling the connectivity, the reflecting elements may be mutually coupled as they are typically separated by less than a wavelength, which may give rise to electrical \emph{mutual  coupling} \cite{gradoni2021end,shen2020modeling}. 
%%The impact of mutual coupling can be modeled via the so-called \textit{mutual impedance matrix} $\bS$, 
%Note that the electrical coupling effect is completely fixed post manufacturing. In contrast, while  the structure of the scattering matrix $\mathbf{\Theta}$ is also determined during manufacturing, i.e., the SC, PC, or FC architectures, $\mathbf{\Theta}$ is tunable via reconfigurable effective impedance matrices $\bZ$ representing the intelligence of IRSs.  

\begin{table}[t]
  \centering
  \caption{Comparison of different models and properties of TC framework}
    \begin{tabular}{|c|c|c|l|c|}
    \hline
           & \textbf{IDS Model} & \textbf{PHY Model} & \multicolumn{1}{c|}{\textbf{INW Model}} \\\hline
     \textbf{Modeling}& Phase shift matrix  $\mathbf{\Phi}$   & GRCS matrix $\mathbf{G}$   & \multicolumn{1}{c|}{Scattering matrix $\mathbf{\Theta}$}  \\\cline{1-4}
         & Diagonal & \multirow{3}{*}{\shortstack{Entries generated\\by $g(\mathbf{\Psi}_\mathrm{t},\mathbf{\Psi}_\mathrm{r})$}}    & \multicolumn{1}{l|}{Single-connected: $\mathbf{\Theta=\Phi}$}       \\
     \textbf{Properties}     &  \multirow{2}{*}{\shortstack{Unit modulus entries \\ in form of $e^{\jmath\theta} $}}   &       & \multicolumn{1}{l|}{Fully-connected: Complex symmetric unitary}   \\
          &        &       & \multicolumn{1}{l|}{Partially-connected: Block diagonal}     \\\hline

    \multirow{2}{*}{\textbf{Advantages}} & \multirow{2}{*}{\shortstack{Accounts for basic\\ IRS properties}} &  \multirow{2}[1]{*}{\shortstack{For large IRSs\\with AoA \&\\AoD modeling} }      & \multicolumn{1}{l|}{Magnitude and phase adjustment of EM waves}     \\
        &      &      & \multicolumn{1}{l|}{Performance improvement}    \\\hline
   
    \multirow{2}{*}{\textbf{Limitations}} & Non-physical model     & \multirow{2}{*}{Only for far-field} & \multirow{2}{*}{Higher hardware complexity}  \\
          & Not scalable     &       &       \\
    \hhline{|=|=|=|=|}
    &\multicolumn{3}{c|}{\textbf{TC Framework}} \\\hline
    \textbf{Compatibility}&\checkmark\checkmark&\checkmark\checkmark\checkmark&\multicolumn{1}{c|}{\checkmark}\\\hline
    \textbf{Advantages}&\multicolumn{3}{c|}{Highly scalable} \\\hline
    \textbf{Limitations}&\multicolumn{3}{c|}{Codebook-dependent} \\\hline
    \end{tabular}%
  \label{tab:addlabel}%
\end{table}%

\subsection{Tile and Codebook-based (TC) Framework}
\label{Sec:TCmodel}

For large IRSs, optimizing each individual reflecting element and estimating the corresponding channel gain may be infeasible in practice. 
To address this issue, a framework for scalable IRS optimization was proposed in \cite{najafi2020intelligent} which relies on the following two design concepts:
\begin{itemize}
	\item The IRS reflecting elements are divided into $N$ subsets, referred to as \textit{tiles}.
	\item Instead of individually configuring each reflecting element, a predefined set of $M$  phase shift configurations for all reflecting elements of a given tile, referred to as \textit{transmission modes},  are designed in an offline stage and stored in a codebook.
\end{itemize}   
Under this framework, for online transmission or  channel estimation, a suitable IRS transmission mode is selected from the codebook. The TC framework can be applied to the IDS, PHY, and INW IRS models, e.g., see \cite{najafi2020intelligent} for the combination of the TC framework and the PHY model.   

When each tile comprises only one reflecting element (i.e., $N$ is equal to the number of reflecting elements), the TC framework reduces to the conventional non-TC framework that does not enable scalable IRS design. 
The other extreme case is that the entire IRS is one tile (i.e., $N=1$), which implies that  a large number of transmission modes $M$ have to be included in a high-dimensional codebook to achieve satisfactory communication performance. 
Therefore, both $N$ and $M$ should not be chosen exceedingly large to
%facilitate IRS model scalability.
%The ideal case is that each tile comprises many reflecting elements ($N$ is not large) while the phase configuration of each tile switches between only a small number of modes ($M$ is small) enabling, e.g., anomalous reflection and absorption. 
%Meanwhile, the joint optimization of all tiles facilitates more sophisticated IRS capabilities, e.g., beam focusing. 
%The ideal case is, while each tile comprises many reflecting elements, the codebook is defined by a series of simple functionalities  (e.g., anomalous reflection), meanwhile, the joint optimization of all tiles enables more sophisticated  IRS capabilities (e.g., beam focusing). 
%In order to 
strike a good balance between  scalability and  achievable performance,  which shall also be illustrated via a case study in Section IV.

\section{Design Challenges and Solutions}
In this section, we identify several key challenges for the design of IRS-assisted wireless systems and provide potential solutions.

\subsection{Joint Design of Active and Passive Beamforming}
To realize the performance gains promised by IRSs, the transmit beams have to be delicately shaped via both the active antennas at the transmitter (Tx) and the \emph{passive} IRS reflecting elements. 
However, the resulting joint active and passive beamforming algorithm design problem gives rise to new technical challenges.

\begin{itemize}
	\item \textbf{Multiplicative optimization variables}: Since IRSs are a part of the wireless channel, the passive beamforming matrix at the IRS is naturally multiplied with the conventional active beamforming vectors. 
	As a result, the joint active and passive beamforming design leads to an intrinsically challenging non-convex problem. 
	To tackle the multiplication of beamformers, a widely-adopted approach is alternating optimization (AO) \cite{9133130,9183907}. 
	In particular, by dividing the multiplied active and passive beamformers into disjoint blocks, each subproblem associated with a single block is solved alternately. 
	Another approach for handling the multiplication of different beamformers is bilinear transformation (BT) \cite{6698281}. 
	Specifically, BT fundamentally circumvents the multiplication issue by regarding the product of the active and passive beamformers as a new entirety. 
	To guarantee the equivalence of such BT, two additional constraints, namely, a positive semidefinite constraint and a constraint in form of a difference of convex functions, are enforced.
	Subsequently, the transformed optimization problem is solved with the new entirety and constraints while the active and passive beamformers can be accordingly recovered, respectively.
	\begin{table*}[t]
		\caption{Comparison of different technical challenges and potential algorithms}
		\label{tab:modelcomparison}\footnotesize
		\newcommand{\tabincell}[2]{\begin{tabular}{@{}#1@{}}#2\end{tabular}}
		\centering
		\begin{tabular}{|c|c|c|}\hline
			\hspace*{0mm} \textbf{} & \textbf{Technical challenge} & \textbf{Algorithm}\\
			\hline
			All models and framework& Multiplicative variables&AO, BT\\\hline
			IDS model & Unit modulus constraint& MO, IA, SCA\\
			\hline
			PHY model & Binary and unit modulus constraint& BnB, MO\\
			\hline
			INW model & Complex symmetric unitary constraint & MO\\
			\hline
			TC framework & Binary constraint& BnB, QP, ADMM\\
			\hline
		\end{tabular}
	\end{table*}
	\item \textbf{IRS-induced constraints}: 
	As mentioned in Section II, the modeling of the IRS itself leads to different constraints for beamforming optimization algorithm design. 

	\hspace{1em}
	\emph{IDS model:} For the IDS model, each diagonal element of the phase shift matrix $\mathbf{\Phi}$ is forced to admit a unit modulus.
%	To tackle this highly non-convex constraint, a series of suboptimal approaches have been developed.
	Since the resulting unit modulus constraint defines a complex circle manifold, one may resort to the application of manifold optimization (MO) theory \cite{9322519}. 
	Alternatively, the unit modulus constrained problem can be equivalently transformed to a rank-constrained problem, which can be further rewritten as a constraint in  form of  a difference of matrix norm functions. This facilitates the design of tractable algorithms by adopting inner approximation (IA) and successive convex approximation (SCA) techniques \cite{9133130}. 
	
	\hspace{1em}
	\emph{PHY model:} The optimization of the GRCS in the PHY model involves in general a combination of binary programming for the selection of reflection beams and a unit-modulus optimization for determining the wave-front phase of each beam \cite{najafi2020intelligent}. 
	Such problems can be solved by leveraging MO and enumeration-based algorithms, e.g., branch-and-bound (BnB).
	
	\hspace{1em}
	\emph{INW model:} The INW model, although sidestepping the unit modulus constraint, does impose a complex symmetric unitary matrix constraint for the IRS scattering matrix $\mathbf{\Theta}$ \cite{9239335}. As the constraint defines a complex Stiefel manifold, we can tackle this difficulty by resorting again to MO methods.
	
	\hspace{1em}
	\emph{TC framework:} The TC framework introduces binary constraints for transmission mode selection from the codebook, which leads to a mix-integer optimization problem that can be optimally solved by BnB. 
	Besides, a suboptimal solution can be obtained by employing the quadratic penalty (QP) method or alternating direction method of multipliers (ADMM) \cite{WCNC}. 
	
	\hspace{1em}In Table \ref{tab:modelcomparison}, we summarize the constraints introduced by the different models and the TC framework along with some available algorithms for resource allocation design in IRS-assisted wireless systems.
\end{itemize}
\subsection{Channel State Information (CSI) Acquisition}
%The significant performance enhancements benefited from IRS rely on perfect channel state information (CSI)
Accurate CSI is of great importance for the design of IRS-aided systems.
% In general, there are two categories of CSI acquisition approaches. (XX)
Since RF chains are not available at the passive IRSs, it is not possible to estimate the IRS-assisted channels directly by having the IRS emit pilot symbols.
Therefore, novel CSI acquisition methods are required and system design methodologies accounting for the inevitable CSI estimation error have to be investigated \cite{9130088}.
\begin{itemize}
	\item \textbf{Channel estimation}: For the IDS and INW IRS models, 
%	one may estimate the channels by controlling the IRS in an on-off manner \cite{9130088}, where only one IRS element is switched on in each time slot such that its reflected channel can be estimated without interference.
%	In particular, the IRS is switched off to estimate the direct channel between the transmitter (Tx) and receiver (Rx).
%	almost all papers adopt DFT matrix….
%	To further improve estimation performance, discrete Fourier transform (DFT)-based passive beamforming has been widely adopted for configuring IRSs in the CSI acquisition phase, where all IRS elements are switched on  \cite{9322519}.
%	To efficiently estimate the high dimensional cascaded channel, the entire estimation period is typically divided into several phases and one channel is estimated per phase. 
%	For instance, the direct channel and cascaded reflecting channels can be successively estimated by employing a linear minimum mean-squared error estimator in a multi-timescale protocol \cite{9130088}. 
	discrete Fourier transform (DFT)-based passive beamforming has been widely adopted at IRSs for the CSI acquisition of the cascaded channel when the receivers (Rx) are single-antenna devices.
	Yet, when the Rx are equipped with multiple antennas, it is challenging to construct the cascaded channel for CSI acquisition. 
	Accordingly, one can estimate the two segments of the cascaded channels in an AO fashion \cite{9322519}.
	Particular attention may be paid to the PHY model, where the sparsity in the angular domain and propagation paths can be exploited.
	In particular, abundant estimation methodologies can be borrowed from the compressed sensing literature  where  sparsity is leveraged for recovering the channel matrices from the received signals.
	In addition, the CSI acquisition overhead for  algorithms developed based on the TC framework scales only with the numbers of tiles, $N$, and transmission modes, $M$, which are design parameters and can be chosen to trade performance with complexity and/or signaling overhead \cite{jamali2020power}. 
	
	\item \textbf{System design with CSI uncertainty}: 
The design of practical IRS-assisted systems has to be robust against CSI errors. In general, there are two models for characterizing CSI uncertainty, namely, the deterministic CSI error model and the statistical CSI error model. The deterministic model assumes that the CSI error lies in an uncertainty region with a known bound, which leads to infinitely many constraints. A commonly-adopted method is to transform these constraints into a set of linear matrix inequalities by employing the S-procedure. On the other hand, the statistical model assumes that the CSI error follows a complex Gaussian distribution with zero mean and known variance, which results in probabilistic chance constraints. In this case, by investigating the channel distribution and exploiting the corresponding inverse cumulative distribution function, the probability constraints can be replaced by more tractable constraints. Alternatively, one can resort to Bernstein-type inequalities to obtain a safe approximation. However, since the variables appear in product form, as discussed in Section III-A, these techniques are not always directly applicable for IRS-assisted system design. As a compromise, one may exploit suitable inequalities, e.g., the triangle inequality, to decouple the product terms in the intractable constraints, which facilitates the reformulation to a convex problem \cite{9183907}.
\end{itemize}
%\newpage
\subsection{Hardware Impairments}
In practice, hardware impairments of all components of a communication system such as power amplifiers, mixers, analog-to-digital converters, and oscillators, are inevitably non-negligible.
In IRS-assisted wireless systems, hardware impairments mainly arise from two parts:
\begin{itemize}
	\item \textbf{RF chain impairments at Tx and Rx:} One widely-adopted model to characterize the hardware impairments at transceivers is the extended error vector magnitude (EEVM) model \cite{9239335}. 
	A distortion noise is  added to the transmit/received signals to model the hardware impairments of the RF chains of the transceivers. This noise is assumed to be Gaussian distributed with its variance proportional to the power of the transmit/received signals.
	\item \textbf{IRS impairments:} There are two approaches for modeling IRS impairments. First, one may model the reflecting elements as finite-resolution phase shifters. In practice, phase shifters are implemented by positive intrinsic-negative (PIN) diodes and $K$ diodes can provide $2^K$ different phase shift levels.
	Second, similar to RF chain impairments, a phase error term can be added to each IRS reflecting phase shift, which is typically modeled by a uniformly distributed or Von Mises distributed random variable \cite{9239335}. 
	The resulting distortion distribution of each single reflecting element for the IDS model and the phase shift configuration for the PHY  model can be correspondingly derived. 
	However, for the INW model, where the reflecting elements are connected with each other, the effects of finite-resolution phase shifters and the distributions of the total phase distortions cannot be straightforwardly determined.
	Thus, for the INW model, more research is needed to characterize the impact of impairments.
\end{itemize}

Based on the discussions above, the design of IRS-assisted wireless systems considering hardware impairments is rather challenging.
In particular, even for a simple point-to-point transmission, the beamformer vector and IRS reflection matrix appear in both the numerator and denominator of the signal-to-noise ratio (SNR) expression.
Thus, majorization minimization (MM) techniques are effective for optimizing impairment-aware IRS-assisted systems \cite{9239335}. 
Specifically, an effective surrogate function needs to be constructed for the SNR expression in quotient form, such that the optimum is easy to find. 
Intuitively, it can be expected that the SNR will saturate when the transmit power is exceedingly large, even for the optimal design, which is a key difference compared to the case when ideal hardware is available.

\subsection{Multi-IRS Systems and IRS Deployment}
Deploying multiple IRSs in wireless systems is a promising solution to fill possible coverage holes. 
In practice, IRSs are usually installed at fixed locations, e.g., facades of infrastructures.
Therefore, the locations of IRSs should be determined in an one-off manner by exploiting  statistical information of the channels, building distribution, and population density.
Intuitively, it is beneficial to create line-of-sight (LoS) links between IRSs and transceivers to reduce the path loss.
However, the pure LoS channel matrix is generally rank-deficient, which is a major disadvantage for exploiting the multiple-input multiple-output (MIMO) spatial multiplexing gain.
Hence, ideally, multiple physically separated IRSs should be deployed such that they can construct full-rank MIMO channels yet with low path loss.
A promising solution for multi-IRS deployment is to leverage radio maps that capture the long-term statistical information of the radio environment \cite{emil}.

In fact, jointly optimizing multiple IRSs and the other elements of a communication system seems to be a  difficult task at first sight.
Nevertheless, it was revealed in \cite{9133130} that incorporating multiple IRSs does not incur additional difficulties for system design.
First, as the path loss after multiple reflections is huge, reflections between IRSs are negligible.
Besides, the distributed IRSs can be thought of as one virtual ``mega IRS''.
Correspondingly, the IRS reflecting matrices can be stacked and be treated as one optimization variable that captures the impact of all IRS reflections \cite{9133130}.
Similarly, the direct and reflecting channel matrices can also be jointly treated as one effective channel matrix for further optimization.
In this sense, all optimization techniques discussed in this section can be extended to tackling multi-IRS scenarios.
One may also apply the TC framework to reduce the design complexity of the virtual ``mega IRS''.

\section{Case Studies}
In this section, we present two case studies to illustrate the design of  IRS-assisted wireless systems with different design objectives and for different  IRS models.
In particular, we first consider the design of a multi-IRS-assisted secure wireless system under the IDS and INW models, respectively, where CSI uncertainty is taken into account for the joint design of beamforming and artificial noise (AN).
Then, based on the PHY model and the TC framework, an efficient design of a SWIPT system with large-scale IRSs  is investigated.  
\subsection{Secure Wireless Communications via IRSs}
%Multiple IRSs can also be deployed for improving physical layer security in wireless networks.
We consider an IRS-assisted secure communication system that consists of one Tx and multiple legitimate Rx in the presence of potential eavesdroppers \cite{9133130}.
Multiple IRSs are deployed for improving the physical layer security of the wireless network.
%The worst-case scenario is investigated where  both the Tx and the eavesdroppers are equipped with multiple antennas while the legitimate users are single-antenna devices.
%As the eavesdroppers typically hide themselves from the Tx, the acquired CSI of the related channels is coarse.
To characterize the CSI uncertainty of the eavesdropping channels, we adopt the deterministic model discussed in Section III-B.
In this case study, we aim to maximize the system sum-rate while mitigating the information leakage to the potential eavesdroppers by injecting AN.
In particular, we employ AO to optimize the IRS phase shift matrix, the transmit beamforming vectors, and the AN covariance matrix in an alternating manner. 
In addition, the generalized S-procedure is applied to design a robust resource allocation algorithm under CSI uncertainty.
The unit modulus constraint induced by the IDS model is handled by the IA approach while the complex symmetric unitary constraint originated from the INW model is tackled by MO. Finally, the non-convexity of the objective function is overcome by SCA.

\begin{figure}[t]\centering
	\includegraphics[height=6.5cm]{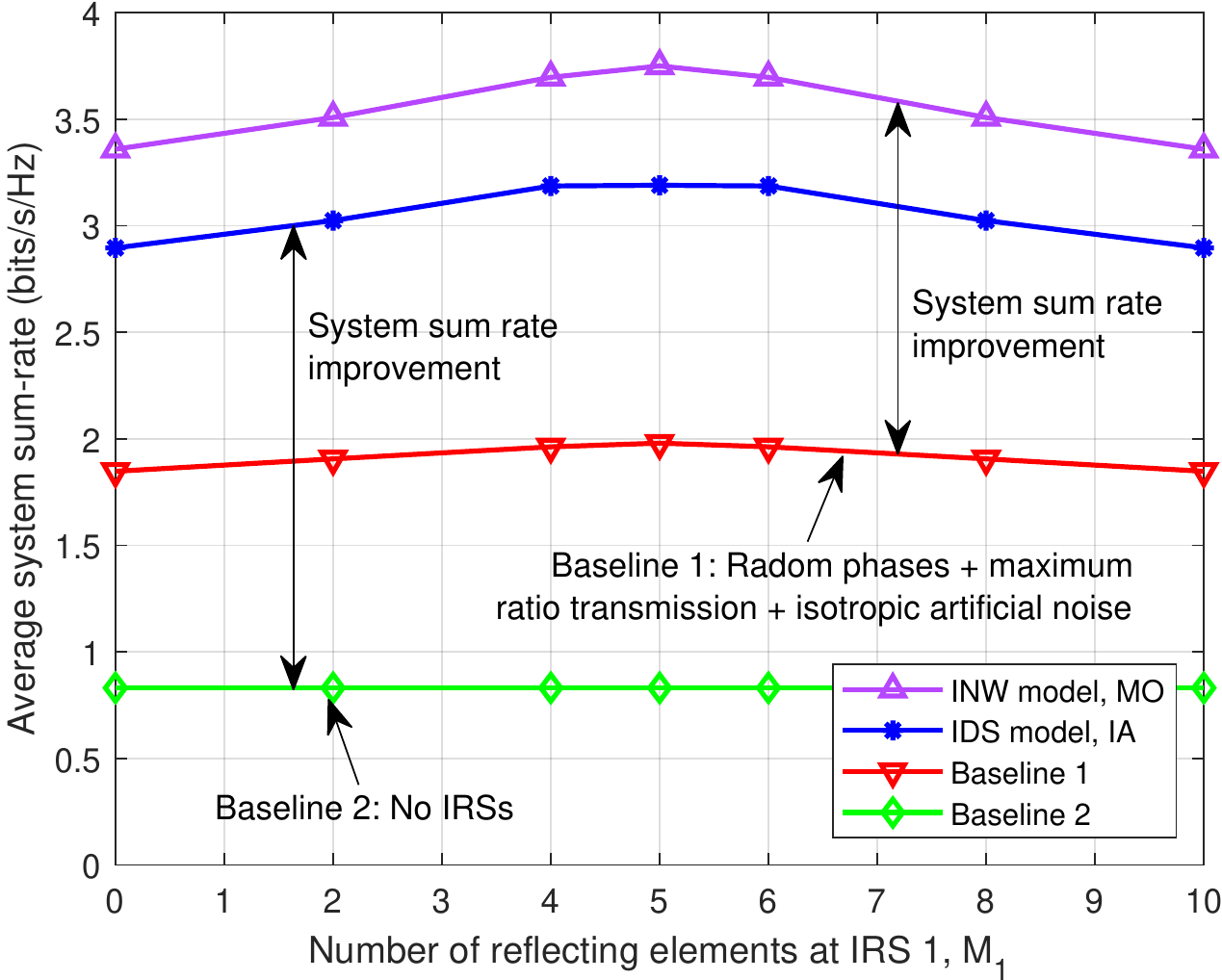}\caption{Average system sum-rate versus the number of reflecting elements at one of the two deployed IRSs. The Tx is assumed to be equipped with 6 transmit antennas. Three single-antenna legitimate Rx and three two-antenna potential eavesdroppers are uniformly distributed in a cell with radius 100 m.
		The total number of reflecting elements of both IRSs is ten.}
	\label{fig6}
\end{figure}

Fig. \ref{fig6} compares the average system sum-rates achieved by deploying a single IRS and two IRSs in a secure wireless network.
%In particular, the Tx is located at the center of a cell while the legitimate users and potential eavesdroppers are uniformly distributed in the cell.
Assume that in total ten reflecting elements are deployed at the IRSs to enhance the communication performance of legitimate Rx that would otherwise be blocked.
The x-axis of Fig. \ref{fig6} represents the number of reflecting elements employed at one of the two deployed IRSs, denoted by $M_1$.
First, we note that the proposed optimized scheme significantly improves the system sum-rate compared to two baselines where a simple transmission technique and no IRSs are employed, respectively.
%Se, we observe that it is beneficial to deploy two IRSs compared with only one (when $M_1=0$ or $10$).
Furthermore, we observe that uniformly distributing the reflecting elements among multiple IRSs ($M_1=5$) is preferable over deploying them at a single IRS (i.e., $M_1=0$ or $M_1=10$) in terms of improving the physical layer security.
This is because multiple IRSs create multiple independent propagation paths which introduce rich macro diversity, and thus, facilitate the establishment of strong end-to-end LoS channels from the Tx to the legitimate Rx, whereas a uniform allocation of reflecting elements can exploit the macro diversity gains more effectively.
Finally, because of the additional DoFs introduced by the impedance network, the average system sum-rate achieved with the INW model is higher than that with the IDS model.  

\subsection{IRS-assisted SWIPT Systems}
Comprising energy-efficient and programmable phase shift elements, IRSs can benefit energy-constrained systems, e.g., SWIPT systems, to provide sustainable high data-rate communication services. 
%On the one hand, the IRS can reflect more power to the desired receivers via the reflecting path, which produces a power gain. On the other hand, when equipped with energy harvesting circuits, the IRS can also harvest radio frequency energy, which enables self-sustainable operation.
Next, to unveil the performance enhancement enabled by employing IRSs in SWIPT systems, we consider a large-scale IRS with $200$ phase shift elements, which can be optimized by invoking the TC framework. 
Moreover, to account for the physical properties of the large IRS, we adopt the PHY model.
For a given transmission mode set generated in an offline design stage, the total transmit power is minimized by jointly optimizing the beamforming at the Tx and the transmission mode selection policy taking into account the quality-of-service requirements of  information decoding receivers and energy harvesting receivers. 
As discussed in Section III-A, we employ a BnB-based algorithm and an SCA-based algorithm to obtain optimal and suboptimal solutions of the formulated mixed-integer optimization problem, respectively.
%%%%%%%%%%%%%%%%%%%%%%%%%%%%%%%%%%%%%%%%%%%%%%%%%%%%%%%%%%%
\begin{figure}[t]
	\centering
	\includegraphics[height=6.5cm]{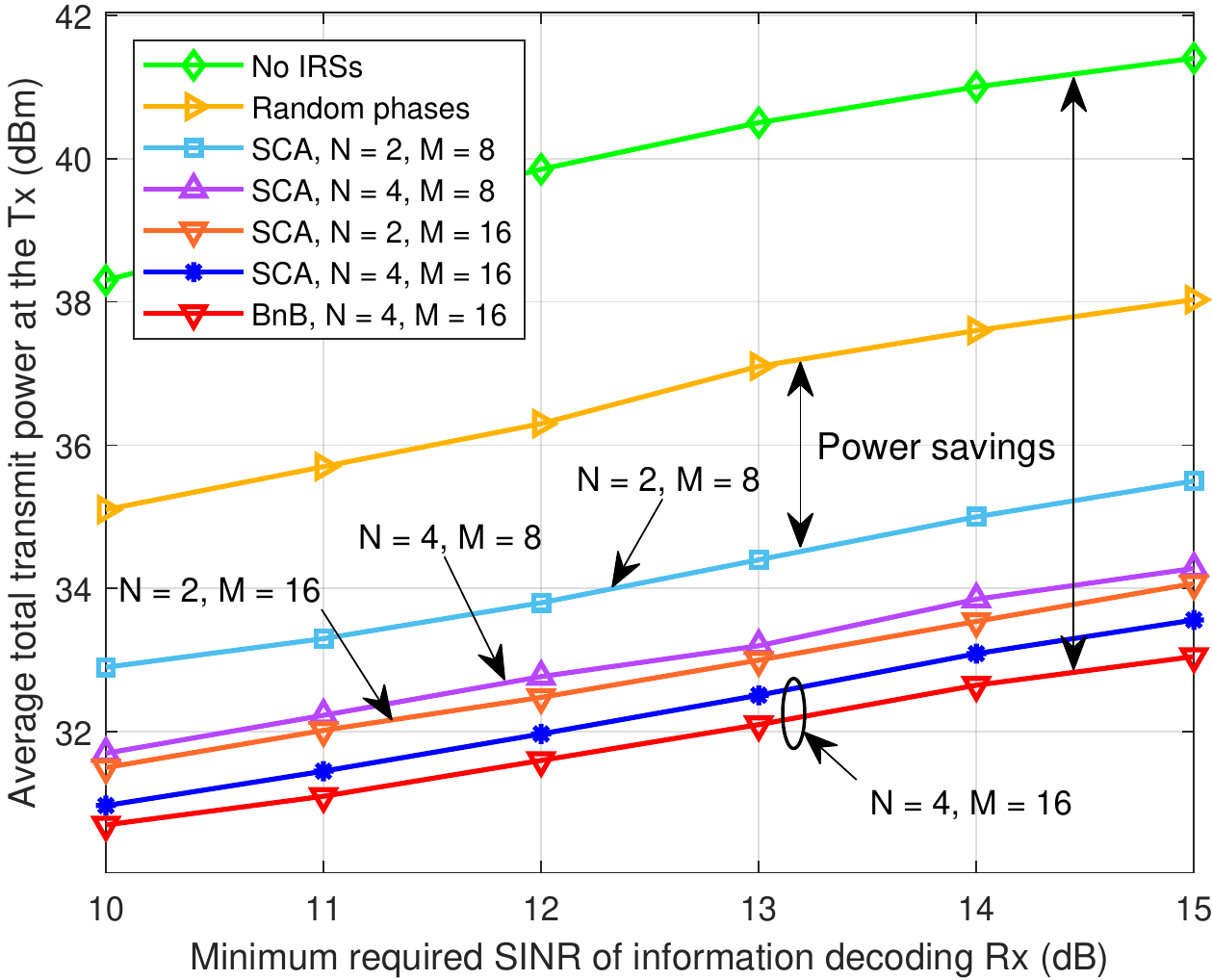}
	\caption{Average total transmit power versus the minimum required SINR of information decoding receivers.
	The Tx is equipped with 6
antennas while the two information decoding receivers and two energy harvesting receivers are single-antenna devices. The IRS is equally divided into $N$ tiles and the size of the transmission mode set is $M$.
}\label{powerSINR_SWIPT}
\end{figure}
%%%%%%%%%%%%%%%%%%%%%%%%%%%%%%%%%%%%%%%%%%%%%%%%%%%%%%%%%%%
\par
In Fig. \ref{powerSINR_SWIPT}, we investigate the average total transmit power versus the minimum required signal-to-interference-plus-noise ratio (SINR) of the information decoding receivers. 
%For comparison, we also consider two baseline schemes. 
%In particular, one scheme does not adopt the IRS while the other scheme employs an IRS with random phase shifts. 
As can be observed from Fig. \ref{powerSINR_SWIPT}, 
%the total transmit powers of the proposed optimal and suboptimal schemes and the two baseline schemes grow with the required SINR. 
%This is attributed to the fact that to satisfy a more stringent minimum SINR requirement, the Tx has to transmit with a higher power. 
%Moreover, 
the proposed optimal and suboptimal schemes yield a significant power reduction compared with the two baseline schemes employing random IRS phase shifts and no IRS, respectively, which reveals the effectiveness of the proposed design methodology for large-scale IRSs. 
Also, we observe that the performance gap between the proposed optimal and suboptimal schemes is small, which verifies the effectiveness of the latter. 
%Compared with the results in Fig. \ref{fig6}, this model clearly shows the scalability of IRS-assisted system design  by jointly adopting the PHY and TC framework and the advantages of the corresponding technical solutions.
Note that by employing the PHY model and the TC framework, the computational complexity of IRS optimization scales only with the number of tiles, $N$, and the sizes of the transmission mode set, $M$.
Fig. \ref{powerSINR_SWIPT} demonstrates that the required transmit power can be reduced by increasing $M$ and $N$, at the expense of a higher computational complexity.
This indicates that by adjusting $M$ and $N$, the PHY model and the TC framework allow us to flexibly strike a balance between  computational complexity and  system performance, which facilitates the efficient and scalable design of large IRS-assisted systems \cite{jamali2020power}.

%The scalability of the PHY model and the TC framework is also confirmed in Fig. \ref{powerSINR_SWIPT}. 
%For fair comparison, we adopt AO-based algorithms to optimize two IRS-assisted SWIPT systems, one based on the IDS model and one based on the PHY model under the TC framework.
%For a practical IRS which usually comprises thousands of sub-wavelength phase shift elements \cite{najafi2020intelligent}, the online element-wise IRS optimization design is unquestionably prohibitive. 
%In contrast, by employing the PHY model and the TC framework, the computational complexity of IRS optimization scales only with the number of tiles and the sizes of the transmission mode set, instead of the number of IRS elements, which facilitates the efficient and scalable design of large IRS-assisted systems 

\begin{figure}[t]\centering
	\includegraphics[height=6.5cm]{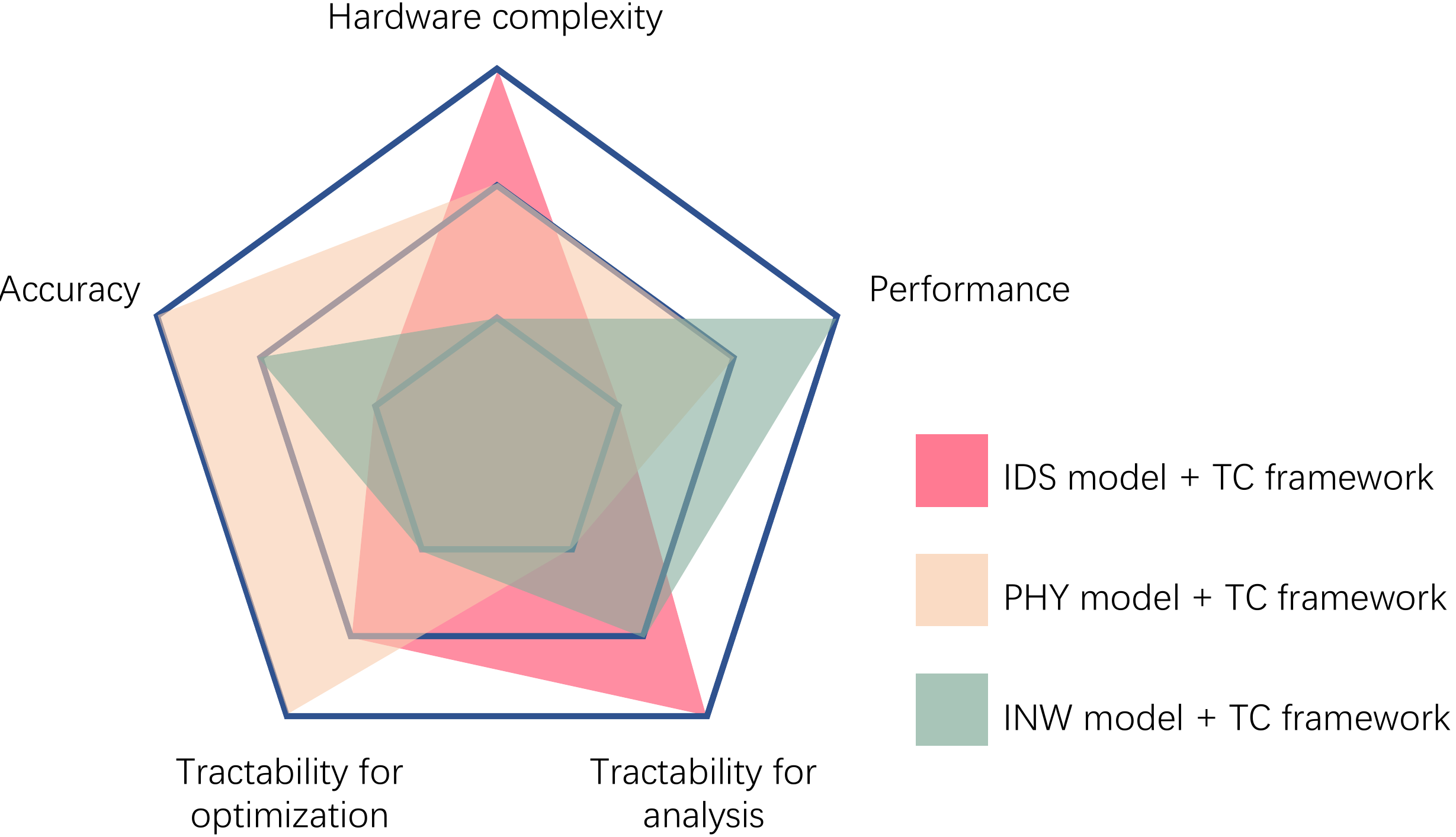}\caption{Comparison of the IDS, PHY, and INW IRS models in terms of hardware complexity, accuracy, tractability for optimization, tractability for analysis, and performance.}
	\label{con}
\end{figure}

\section{Conclusions and Future Research Directions}
In this article, we have provided a comprehensive overview of different IRS models and their implications for the design of IRS-assisted wireless communications systems.
%Special attention has been paid to comparing different IRS models and their impact on the design of IRS-assisted wireless systems.
In particular, thanks to its simplicity, the IDS model has been widely adopted in the literature. To accurately characterize the IRS response to EM waves from different impinging directions, the PHY model was proposed. In addition, at the expense of a higher hardware complexity, the INW model was put forward to allow for connected reflecting elements. Finally, the TC framework was advocated to facilitate the design of large  IRS-empowered systems.
A qualitative comparison of the different IRS models and the TC framework discussed in this article is shown in Fig. \ref{con}.
To unleash the full potential of IRS-enabled wireless communications, there are several open research problems that deserve unremitting efforts.

\textbf{Integrating IRSs into high frequency wireless systems:} High frequency wireless systems, e.g., millimeter-wave and Terahertz communication systems, have received increasing attention in recent years because of the spectrum crunch dilemma. 
However, wireless signals are vulnerable to blockages due to the poor scattering at high operating frequencies.
As such, IRSs are a key enabler to construct an effective virtual LoS link for high frequency communications. 
The PHY model introduced in this article would be an excellent candidate for capturing the properties of limited scattering propagation environments.

\textbf{Design with statistical CSI:} Most design methodologies for IRS-assisted wireless systems rely on instantaneous CSI. 
However, this requires all IRS reflecting elements to be rapidly switched between different phase shift levels, which adds another layer of burden for practical implementation, especially when the channel coherence time is short.
Therefore, designing IRS-aided systems based on long-term  statistical CSI is of great importance to reduce the signaling and hardware implementation complexity.
In addition, while intuitive heuristics have been proposed for IRS deployment, a sophisticated mathematical formulation for IRS position optimization based on long-term CSI is still an open problem.

\textbf{Artificial intelligence-enabled IRS-assisted systems:} Although abundant optimization techniques have been leveraged to design IRS-assisted systems, the resulting computational complexity is still relatively high.
In this sense, artificial intelligence (AI)-based techniques seem promising for the low-complexity design of IRS-empowered systems.
In particular, data-driven deep learning (DL) can be applied to realize truly real-time resource allocation.
On the other hand, model-driven DL exploits explanatory models by exploiting communication domain knowledge and therefore can reduce the demand for huge volumes of training data. 
%In addition to facilitating the efficient design of IRS-assisted systems, AI-based techniques can also exploit the plentiful data to capture the IRS-involved cascaded channel features in a federated learning fashion, which further improves channel estimation accuracy.

% or


%\appendices
%% you can choose not to have a title for an appendix
%% if you want by leaving the argument blank
%\section{}

%\section*{Acknowledgment}

\ifCLASSOPTIONcaptionsoff
  \newpage
\fi



%\bibliographystyle{IEEEtran}
%\begin{thebibliography}{1}
%\end{thebibliography}
\bibliographystyle{IEEEtran}
\bibliography{ref}


%\end{IEEEbiography}
%% insert where needed to balance the two columns on the last page with
%%\newpage


%\vfill


\end{document}